# Effect of Ca doping for Y on structural / microstructural and superconducting properties of $YBa_2Cu_3O_{7-\delta}$

Rajiv Giri[1], H.K. Singh[2], O.N. Srivastava[1], V.P.S. Awana[2,*], Anurag Gupta[2], B.V. Kumaraswamy[2] and H. Kishan[2]

[1] Department of Physics, Banaras Hindu University, Varanasi, U.P. 200005, India.

[2] National Physical Laboratory, K.S. Krishnan Marg, New Delhi 110012, India.

## Abstract

We report systematic studies of structural, microstructural and transport properties of $Y_{1-x}Ca_xBa_2Cu_3O_{7-\delta}$ bulk samples with $0.20 \geq x \geq 0.0$. The partial substitution of $Ca^{2+}$ at $Y^{3+}$ site in $YBa_2Cu_3O_{7-\delta}$ leads towards slightly overdoped regime, which along with disorder in $CuO_2$ planes is responsible for decrease in superconducting transition temperature ($T_c$) of the doped system. The microstructural variants being studied by transmission electron microscopy (TEM) technique in imaging and selected area diffraction modes revealed an increase in the density of twins with increase in Ca concentration despite the fact that there is slight decrease in the orthorhombicity. This is against the general conviction of decrease in twin density with decreasing orthorhombicity. An increase in twin density results in sharpening of twin boundaries with increasing Ca concentration. These sharpened twin boundaries may work as effective pinning centers. A possible correlation between microstructural features and superconducting properties has been put forward and discussed in the present investigation.

**Key words:** $Y_{1-x}Ca_xBa_2Cu_3O_7$ system, Microstructure, Twin boundaries, and $T_c$

* Corresponding Author: e-mail awana@mail.nplindia.ernet.in

**Introduction:**

Ever since the discovery of superconductivity at ~90K in Y bearing cuprates, they are very attractive candidates for investigation due to their varying critical transition temperature ($T_c$) with oxygen content and strong flux pinning capability in high magnetic field and higher irreversibility line. It is now known that $T_c$ is determined by both the hole concentration in the $CuO_2$ planes and relative charge on the oxygen within the planes [1]. The level of this charge can be controlled either by manipulating the oxygen stoichiometry in the Cu-O chain by application of pressure or by atomic substitution [2, 3]. In the present study we focus on atomic substitution of Ca at Y site.

The substitution of non-isovalent cations for the native ones is of great interest, because it is expected that by such substitutions the charge distribution in the structure, and consequently, the concentration of charge carriers could be varied. Therefore, the effect of substitution on exploration of the sensitivity of $T_c$ and other physical properties to the hole concentration in high $T_c$ superconductors has proved to be a viable approach [1]. Hence substitution of $Y^{+3}$ by $Ca^{+2}$ ions with similar ionic radius but lower valence may increase the hole concentration in YBCO system.

Since the valence of $Ca^{2+}$ is smaller than that of $Y^{3+}$, Ca doping will supply the mobile holes into the $CuO_2$ planes and in fact substitution of $Ca^{2+}$ for every $Y^{3+}$ ion increases carrier concentration with ½ hole per plane. Substitution of Ca has been investigated in both deoxygenated [4-6] and oxygenated [7-9] $Y_{1-x}Ca_xBa_2Cu_3O_{7-\delta}$ (where $0\leq \delta \leq 1$) system. It is known that Ca is substituted mostly at the Y position [7-10], although substitution at Ba site may also take place. The Ca is doping in deoxygenate Y-123 compound leads to insulator to metal transition and introduces superconductivity in the compound [12-14]. If the samples are fully oxygenated, Ca substitution causes over-doping of Y-123 compound and $T_c$ decreases [7, 8]. However, this is not always true, in fact it is suggested that Ca substitution induces disorder in adjacent $Cu\text{-}O_2$ planes, which is partly responsible for $T_c$ depression [9-11]. Earlier, Schmehl et. al. have obtained increased grain boundary critical current density ($J_c$) by partial replacement of yttrium in $YBa_2Cu_3O_{7-\delta}$ with calcium [15]. The previous studies indicate that reduction of charge at Y site due to Ca substitution is partially balanced by loss of oxygen [9, 11, 16-18]. Consequently hole concentration may either increase or decrease in comparison to virgin compound depending upon the balance between hole doping by $Y^{3+}/Ca^{2+}$ substitution and the oxygen loss [9,11,16-20].

Although the phase diagram for variable Ca contents in both oxygenated deoxygenated Y-123 system is extensively investigated [4-20], the effect of Ca substitution on microstructural variants and consequently superconducting properties of oxygenated Y-123 compound has not yet been studied so far in detail. The central aim of the present work is to investigate the structural and microstructural changes due to substitution of Ca at Y site and their possible correlation with superconducting properties.

**Experimental**

A series of samples of $Y_{1-x}Ca_xBa_2Cu_3O_{7-\delta}$ (where $0 \leq x \leq 0.2$) have been prepared by standard solid state reactions. The appropriate amount of high purity powders of $Y_2O_3$, CaO, $BaCO_3$ and CuO have been mixed, grounded and calcined at 920 $^0$C for 48 hours in flowing oxygen with two intermediate grindings. The resultant mixture is grounded and pelletized at a pressure of 3.0-tons/ inch$^2$. Subsequently, the samples were sintered in flowing oxygen at 925 $^0$C for 24 hours and cooled slowly to room temperature. All the samples in the present investigation were subjected to gross structural characterization by x-ray diffraction technique (XRD, Philips PW 1710, $CuK_\alpha$), electrical transport measurement using four-probe technique (Kiethley Resistivity-Hall setup), AC susceptibility measurement employing Lakeshore-7000 susceptometer and microstructural investigations by employing a transmission electron microscope (Philips EM-CM-12) in imaging and diffraction modes.

**Results and Discussions:**

The as grown samples having various Ca concentrations were subjected to gross structural characterization employing x-ray diffraction technique (XRD) using Cu $K_\alpha$ radiation. The typical x-ray diffraction patterns of high temperature superconducting $Y_{1-x}Ca_xBa_2Cu_3O_{7-\delta}$ compounds (where $0 \leq x \leq 0.2$) are shown in Fig.1. The XRD patterns reveal that all the samples are polycrystalline and single phase in nature and correspond to orthorhombic Y(Ca):123 phase. The variation of lattice parameters *a*, *b* and *c* with Ca concentration in $Y_{1-x}Ca_xBa_2Cu_3O_{7-\delta}$ is shown in Fig.2. It may be noticed that the lattice constant *a* increases with Ca addition while lattice constant *b* remains almost constant. The *c* lattice parameter does not change significantly with Ca addition.

On the other hand, addition of Ca at Y site decreases slightly the difference between *a* and *b* parameters and thus reduces the orthorhombicity $(a-b)/(a+b)$ of Y:123 phase (as shown in Fig.3). This decrease in orthorhombicity can be visualized with the help of characteristic orthorhombic splitting of reflections (012), (102) and (013), (103) as shown in Fig.4(a) and Fig.4(b) respectively. Merging of (012) and (102) peaks and shifting of (102) reflection towards lower angle with increasing Ca content in fig.4(a) indicates that orthorhombicity of the system decreases with addition of Ca. Similar behavior can also be seen for (013) and (103) reflections [Fig.4(b)].

The temperature dependence of the resistance of the as synthesized $Y_{1-x}Ca_xBa_2Cu_3O_{7-\delta}$ HTSC samples was measured by the standard four-probe technique. In the normal state all the samples show metal like behavior. The Ca free sample has the highest transition temperature ($T_{c(R=0)}$) of 91K. With the substitution of $Ca^{2+}$ for $Y^{3+}$ the critical transition temperature of the samples decreases. The $T_c$ values of the as grown ($Y_{0.95}Ca_{0.05}$ ):123, ($Y_{0.85}Ca_{0.15}$ ):123 and ($Y_{0.80}Ca_{0.20}$ ):123 HTSC phases are 87K, 86K and 85K respectively. The representative R-T behavior of as grown $Y_{1-x}Ca_xBa_2Cu_3O_{7-\delta}$ (where $0\leq x \leq 0.2$) samples are plotted in Fig.5(a). It may be noticed that all the samples, which are subjected to $T_c$ measurement, have similar geometry and accuracy in $T_c$ measurement is ±0.1K. The depression of $T_c$ with Ca addition may be either due to decrease in oxygen content in the CuO chains [16] or due to trapping of mobile holes or some mechanism connected with oxygen vacancy disorder [9-11, 13-15, 17-21]. Also, from our investigations of R-T behaviour, it may be inferred that normal state conductivity increases (or R decreases) with increasing Ca concentration in $Y_{1-x}Ca_xBa_2Cu_3O_{7-\delta}$ . This suggests the occurrence of overdoped regime and, as a result a decrease in $T_c$. The variation of $T_c$ with Ca content is shown in Fig.5(b). The variation of $T_c$ with x in $Y_{1-x}Ca_xBa_2Cu_3O_{7-\delta}$ was also confirmed by AC susceptibility measurements and the obtained results are plotted in Fig. 5(c). Here the $T_c$ is determined by the onset of the diamagnetic signal and is generally called $T_{c(dia)}$. Interestingly, $T_{c(dia)}$ is observed to be lower than $T_{c(R=0)}$, primarily due to the reason that the former is a bulk property of the material and the latter depends only on a percolative / filamentary path. The variation of $T_c$ by $Y^{3+}/Ca^{2+}$ substitution in Y:123 systems with variable oxygen content, which is as such not the central point of our article, had been discussed a lot in earlier studies [4-21]. In fact induction of mobile carriers [5-7], coupled with various structural changes [11], overall oxygen content and the disorder in $Cu\text{-}O_2$ planes [9, 10] are various mechanisms responsible for variation of $T_c$ in theses systems. It is possible that $T_c$ and the number of holes are not directly related to the each other but various other factors also simultaneously affect the $T_c$ [22].

In order to explore possible effect of Ca doping on microstructural characteristics in $Y_{1-x}Ca_xBa_2Cu_3O_{7-\delta}$ cuprate superconductors, the transmission electron microscopic technique was employed in imaging and selected area diffraction (SAD) modes. The detailed structural and micro structural investigations on the Ca free and the Ca doped Y: 123 unravel some new and peculiar features in terms of twinning bands and broadening of the twin boundaries. The TEM micrograph corresponding to the Ca free Y:123 phase is shown in Fig. 6 (a). The presence of wide twin bands and broadened twin boundaries are easily discernible. The pristine Y:123 phase as usual has been found to be orthorhombic (a=3.82 Å, b=3.88 Å and c=11.73 Å) and tends to form defect substructures, e.g., dislocations, twins, etc. due $Y^{3+}/Ca^{2+}$ substitution. The formation of twins is also related to the tetragonal to orthorhombic structural phase transformation. The selected area diffraction (SAD) pattern corresponding to Y:123 is shown in Fig. 6 (b). The elongation and the splitting of the prominent diffraction spots along the [110] direction is indicative of 45º type twins. In the case of 45º twin *a* and *b* axes constitute a mirror image along the [110] direction twin axis. Since a=3.82 Å and b=3.88 Å, the angle subtended a and b axes are 89º and 91º respectively. This also results in bending of the [110] plane across twins ~2º. In addition to the wide twin bands others feature present are dislocations and the broadening of the twin boundaries. The genesis of the broadened twin boundaries may be understood in terms of the high value of the orthorhombicity $\approx 7.9 \times 10^{-3}$ for Ca free Y:123. The representative TEM micrographs of 15% Ca doped Y: 123 are shown in Fig. 6 (c, d). In this case too the microstructure (Fig.6c) reveals presence of twin bands, though the density of the twin bands has increased in this case and in addition the broadening has also been reduced and the bands appear sharper than in case of Ca free Y:123. It may be mentioned that by substitution of Ca for Y in the fully oxygenated Y: 123 the orthorhombicity has been found to decrease, e.g., the orthorhombicity of 15 % Ca doped Y: 123 is $7.6 \times 10^{-3}$ which lower than that of the Ca free Y: 123. Thus it may be conjectured that the reduced orthorhombicity may one of the factors responsible for decrease in the twin boundary broadening. The SAD pattern corresponding to the microstructure of 15% Ca doped Y: 123 (as shown in fig. 6c) is shown in Fig. 6 (d) and the reduced elongation and splitting of main diffraction spots in the [110] direction in this case are indicative of sharper boundaries twin boundaries. Another interesting result of the present investigation is the increased density of the twin bands with increasing Ca concentration and this observed feature is against the general conviction that lower orthorhombicity corresponds to lower density of twins and as orthorhombicity vanishes the twins also disappear as in case tetragonal Y:123[ ]. Thus the increased density of twin bands in Ca doped Y:123 implies that orthorhombicity may not be the singular factor responsible for formation of twin bands. Presumably, another important parameter that may affect the occurrence and density of twins seems to be the 'anisotropy' of the system. In less orthorhombic Ca doped Y:123 compound this may be the dominating factor in twin formation. Hence an increase in the anisotropy corresponds to higher energy state of the system and consequently the Ca doped Y: 123 system will try to lower the free energy by formation of defect structures such as twins. This will therefore lead to an increase in the density of the twin bands in the Ca doped Y: 123 system. This is well evidenced by our TEM investigations which reveal twin density to be 1.25 x $10^5$ $cm^{-2}$ and 2.84 x $10^5$ $cm^{-2}$ for Ca = 0 % and 15 % respectively. In order to verify that with increasing Ca concentration the 'anisotropy' contribution is dominating over the orthorhombic counterpart we have further carried out selected area structural and microstructural probing of the 20 % Ca doped Y:123 and the TEM micrographs are shown in Fig.6 (e,f). As seen in Fig. 6e twin boundaries become further sharper and the twin density also increases and in fact in this case the twin density is 3.75 x $110^5$ $cm^{-2}$. These microstructural features can be explained on the basis of further decrease in orthorhombicity accompanied by a corresponding increase in the anisotropy of the Ca doped Y: 123 system. As seen in the SAD pattern of fig. 6f the much reduced elongation and splitting of the main diffraction spots is also in conformity with the sharper twin boundaries.

Therefore, it can be concluded that sharpening of twin boundaries is related to the decrease in orthorhombicity of Ca doped Y-123 compounds. Thus TEM investigations of as grown $Y_{1-x}Ca_xBa_2Cu_3O_{7-\delta}$ (where $0 \le x \le 0.2$) samples clearly reveal a correlation between the Ca concentration and curious microstructural features in the form of sharpening of twin boundaries with their increased density. Such microstructural features will be beneficial in understanding and tailoring the physical and superconducting properties of Y-123 HTSC, in particular the inter-grain critical current density of these compounds. Very recently [24], substantial enhancement had been observed in irreversibility field and critical current of $Y_{1-x}Ca_xBa_2Cu_3O_{7-\delta}$, and is explained solely on the basis of overdoping. Our present results pertaining to structural and micro-structural details of $Y_{1-x}Ca_xBa_2Cu_3O_{7-\delta}$ system shows that besides overdoping, the sharpened twin boundaries being acting as effective pinning centers might be responsible for reported improved irreversibility line and critical current characteristics.

## Conclusion

In the present study, we have investigated the effect of Ca doping at Y site on the structural /microstructural characteristics and superconducting properties of $YBa_2Cu_3O_{7-\delta}$ HTSC. The superconducting properties in particular the $T_c$, has been found to decrease with increasing doping level of Ca. This decrease in $T_c$ value has been attributed to the overdoping and disorder in $CuO_2$ planes being induced by Ca doping in $YBa_2Cu_3O_{7-\delta}$. XRD results reveal that orthorhombicity of the system decreases with increasing doping level of Ca. Microstructural investigations employing TEM reveal that twin boundary becomes sharper with increasing doping level of Ca in $YBa_2Cu_3O_{7-\delta}$.

## Acknowledgement

Authors would like to thank Prof. A.V. Narlikar Scientist emeritus, IUC Indore for his helpful discussions and valuable comments**.** Authors at NPL are further thankful to DNPL, Prof. Vikram Kumar for his encouragement and keen interest in the present work.

.

**Figure caption**

**Fig. 1.** Representative X-ray diffraction patterns of $Y_{1-x}Ca_xBa_2Cu_3O_{7-\delta}$ with (a) X=0.00 (b) X=0.05 (c) X=0.15 (d) X=0.20 , as synthesized samples using $CuK_\alpha$ radiation.

**Fig. 2.** Variation of 'a', 'b' and 'c' lattice parameters as a function of doping level x in $Y_{1-x}Ca_xBa_2Cu_3O_{7-\delta}$ HTSCs.

**Fig. 3.** Variation of orthorhombocity (a~b)/(a+b) as a function of Ca content in $Y_{1-x}Ca_xBa_2Cu_3O_{7-\delta}$.

**Fig. 4.** Variations in the intensity and position of (a) (012) and (102) reflections, (b) (013) and (103) reflections for $Y_{1-x}Ca_xBa_2Cu_3O_{7-\delta}$ compounds.

**Fig. 5(a)** Resistance vs temperature behaviour of as synthesized $Y_{1-x}Ca_xBa_2Cu_3O_{7-\delta}$

**Fig. 5(b)** Variation in critical transition temperature $T_{c\,(R=0)}$ as a function of Ca content in $Y_{1-x}Ca_xBa_2Cu_3O_{7-\delta}$.

**Fig. 5(c)** AC magnetic susceptibility ($\chi$) versus temperature (T) plots for $Y_{1-x}Ca_xBa_2Cu_3O_{7-\delta}$

**Fig. 6(a)** TEM micrograph showing presence of broad twin boundaries (marked by arrow) and dislocations (marked by D) in $YBa_2Cu_3O_{7-\delta}$ HTSC material.

**Fig. 6(b)** SAD pattern corresponding to the region containing twin boundaries in Fig.6(a). The elongation and splitting of diffraction spots along [110] direction is in conformity with the broadened twin boundaries.

**Fig. 6(c)** Microstructure revealing less broadened twin boundaries for $Y_{0.85}Ca_{0.15}Ba_2Cu_3O_{7-\delta}$ in comparison to Ca free Y123 compound.

**Fig. 6(d).** SAD pattern corresponding to Fig. 6(c) showing comparatively less prominent elongation and splitting of diffraction spots along [110] direction.

**Fig. 6(e).** The representative TEM micrograph for $Y_{0.80}Ca_{0.20}Ba_2Cu_3O_{7-\delta}$ revealing sharp twin boundaries.

**Fig. 6(f).** SAD pattern corresponding to the same region of TEM micrograph of $Y_{0.80}Ca_{0.20}Ba_2Cu_3O_{7-\delta}$. .

Fig. 1 Giri et al.

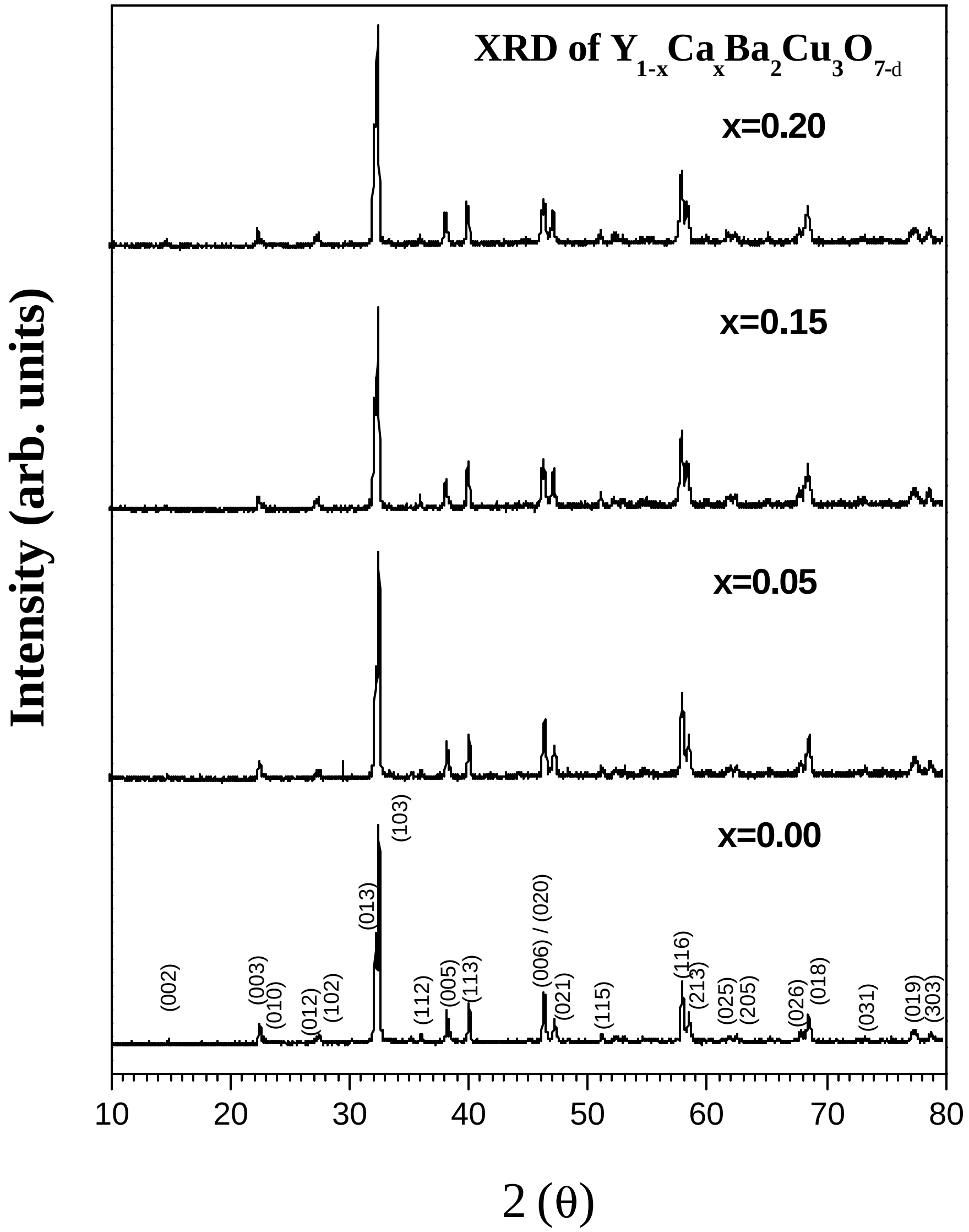

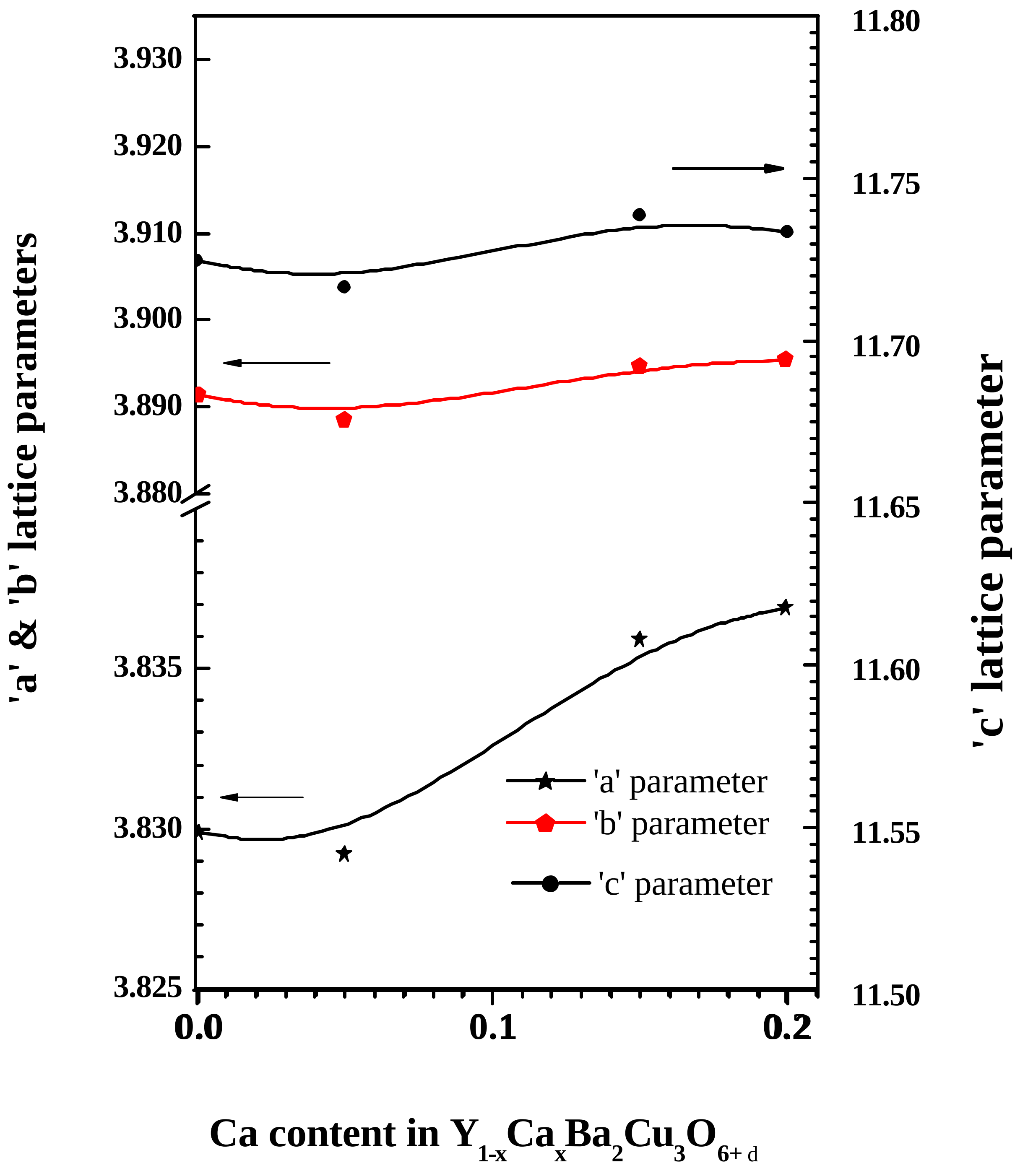

Fig. 3 Giri et al.

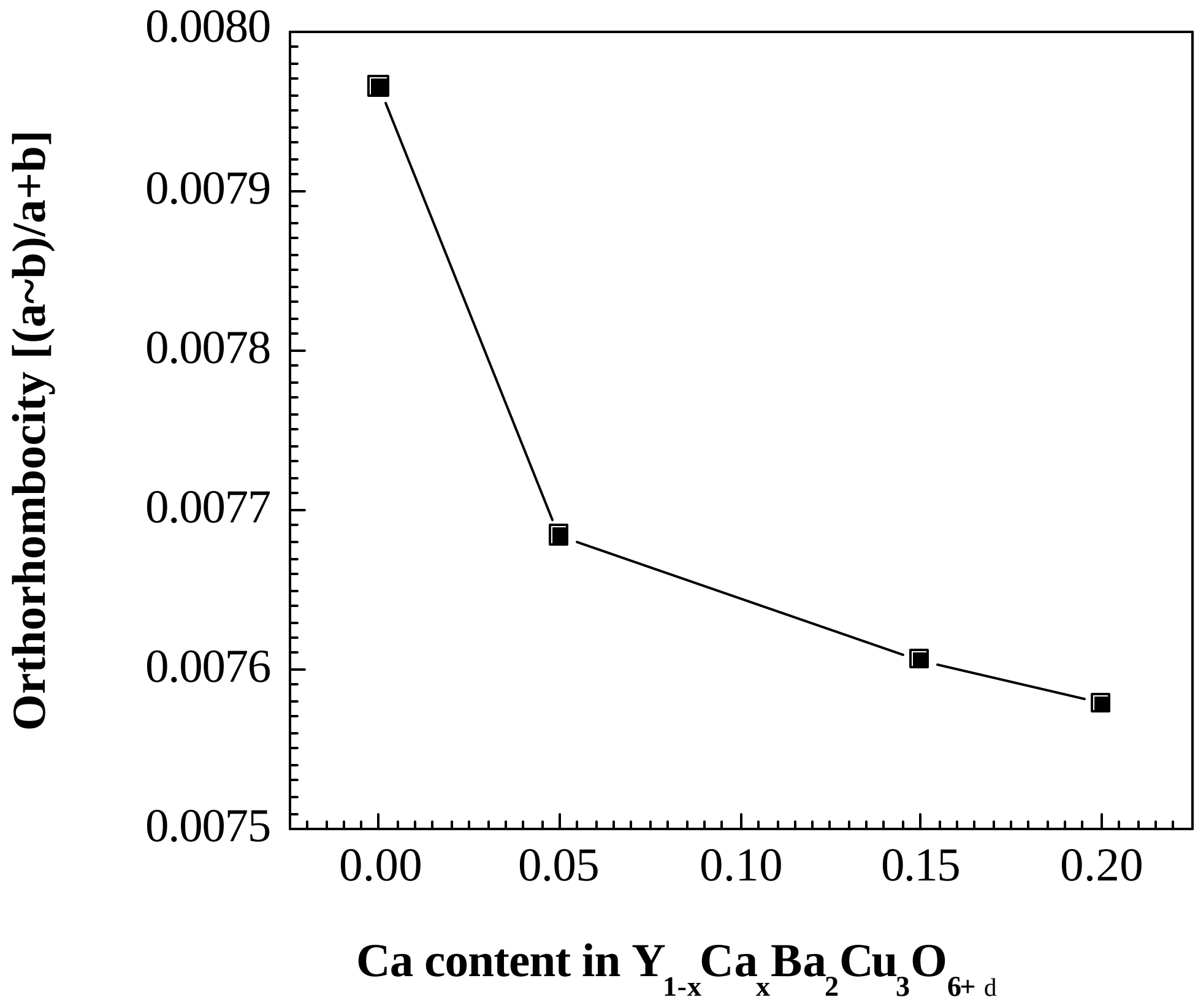

Fig. 4(a) Giri et al.

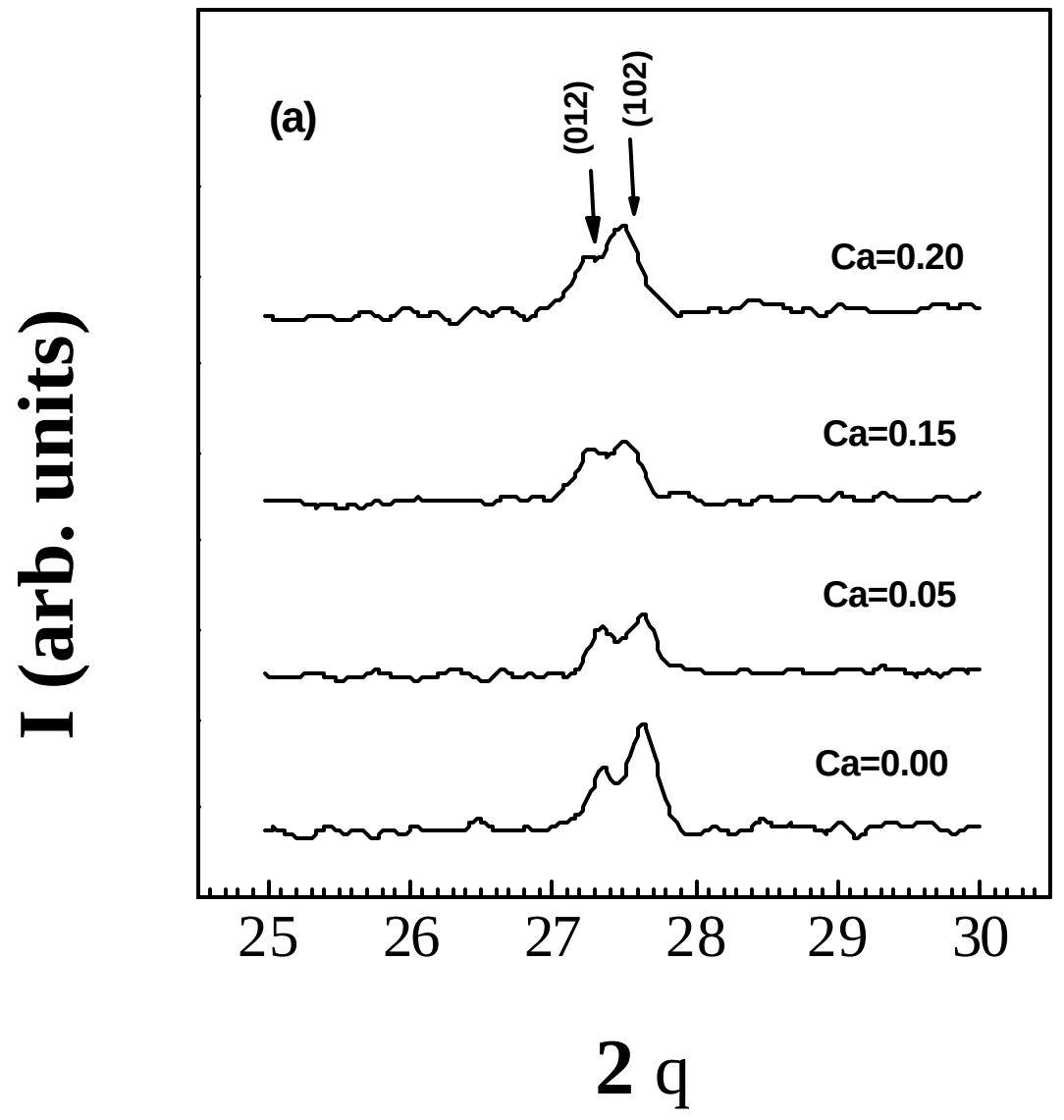

Fig. 4(b) Giri et al.

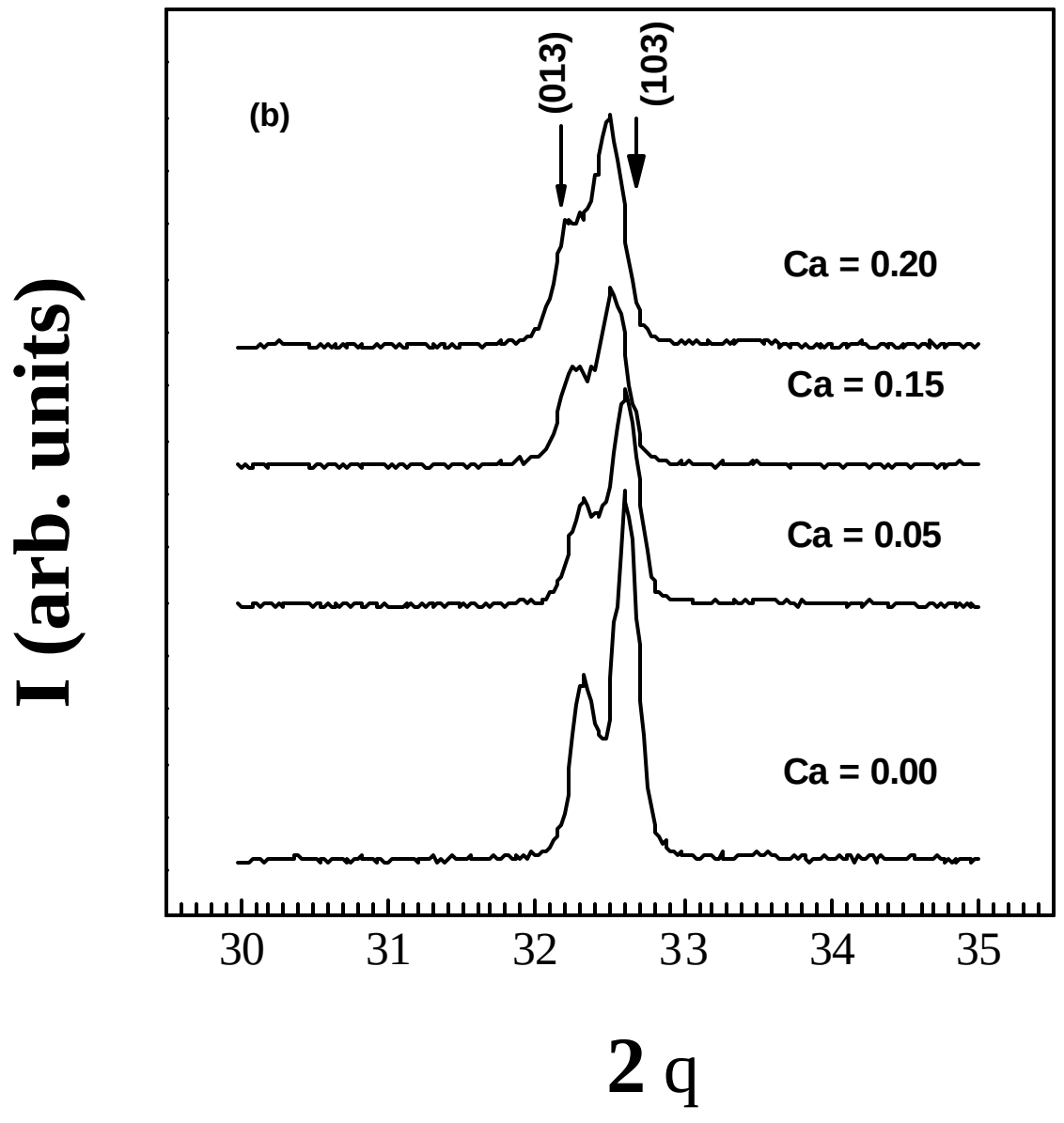

Fig. 5(a) Giri et al.

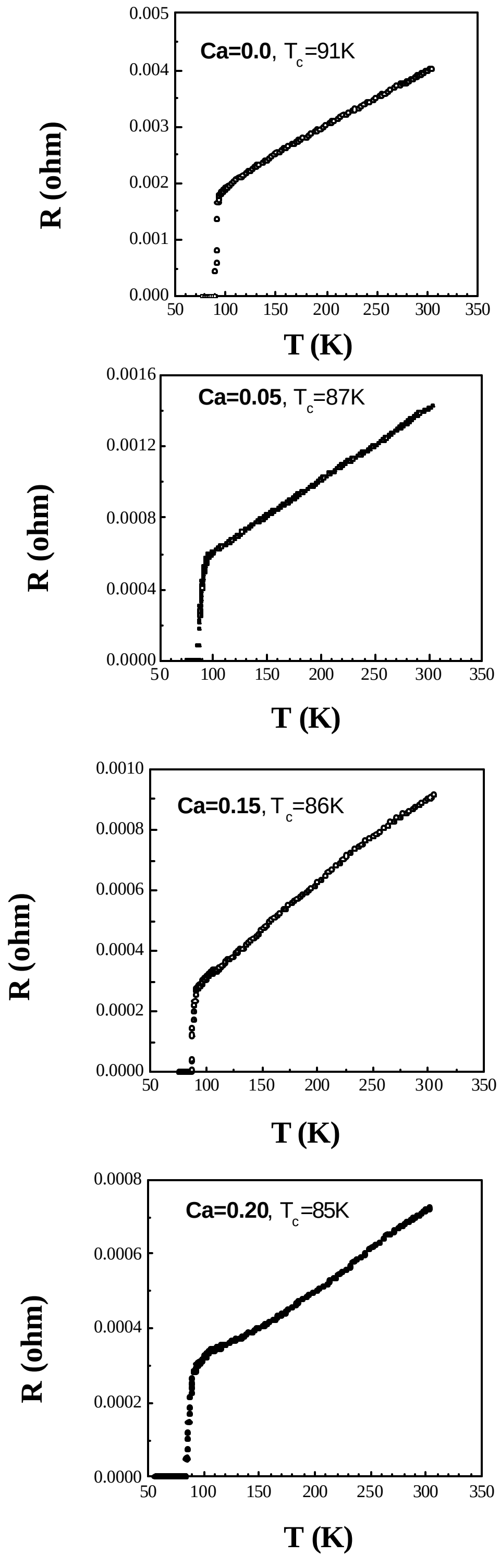

Fig. 5(b) Giri et al.

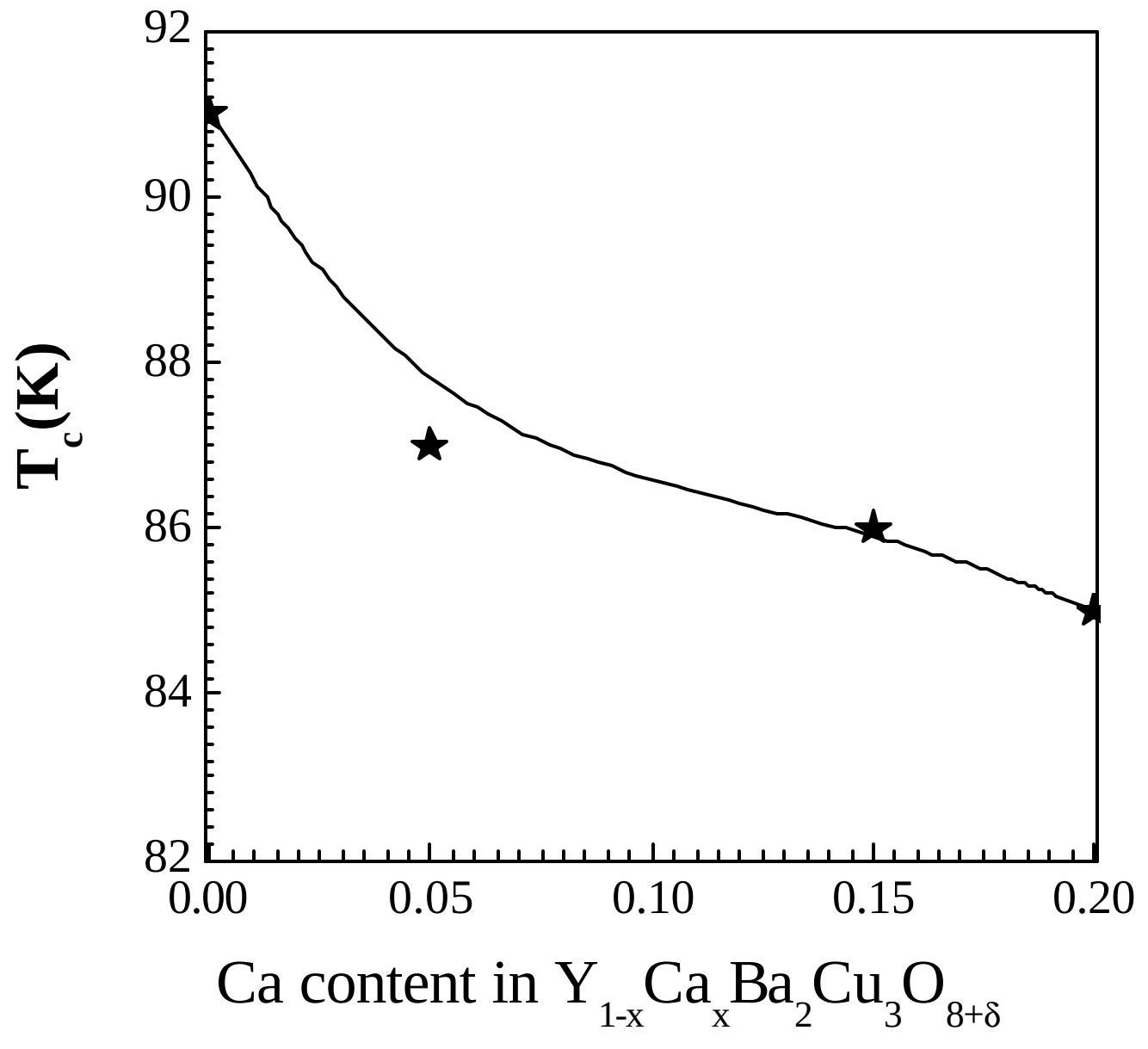

Fig. 5(c) Giri et al.

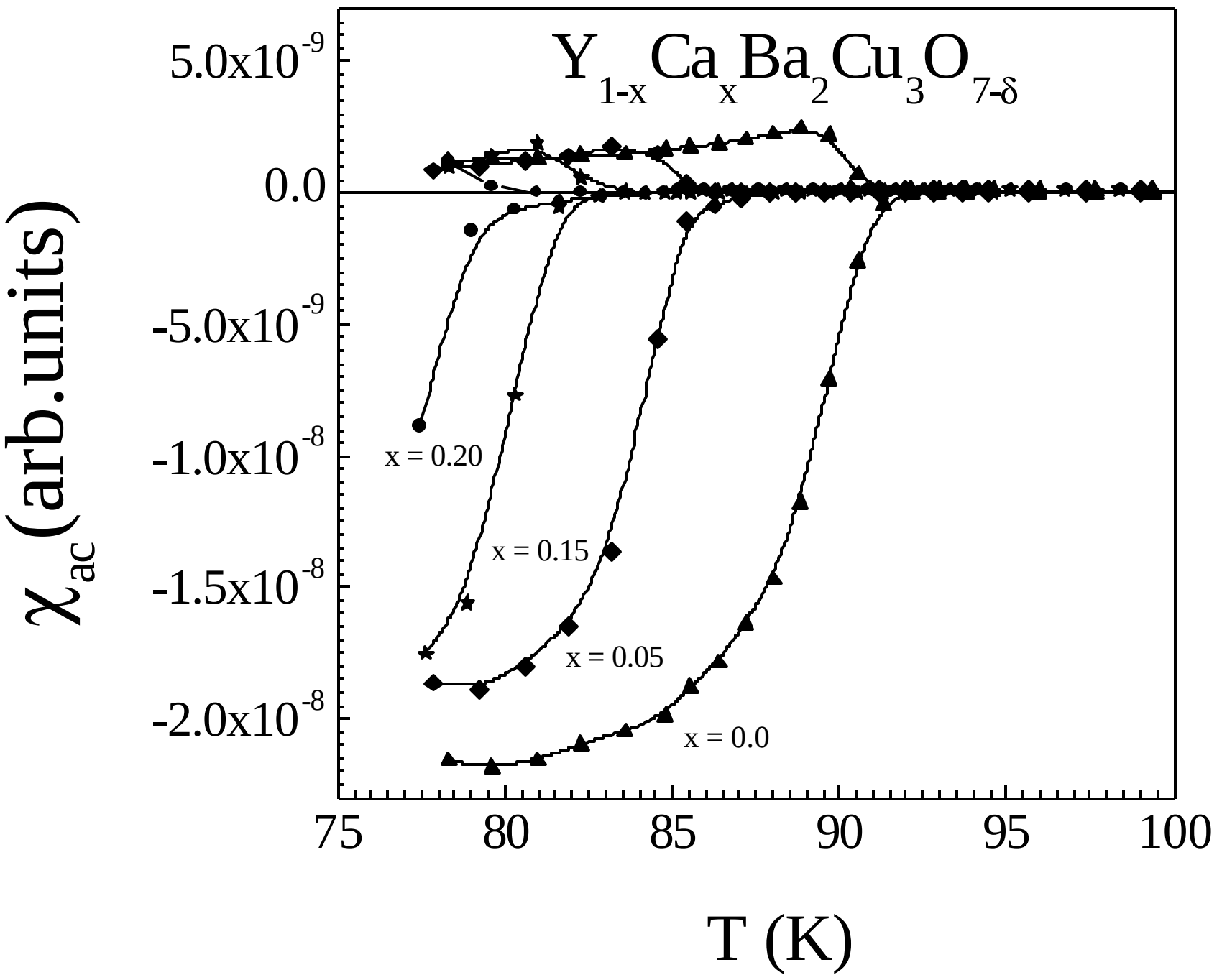

Fig. 6(a) Giri et al.

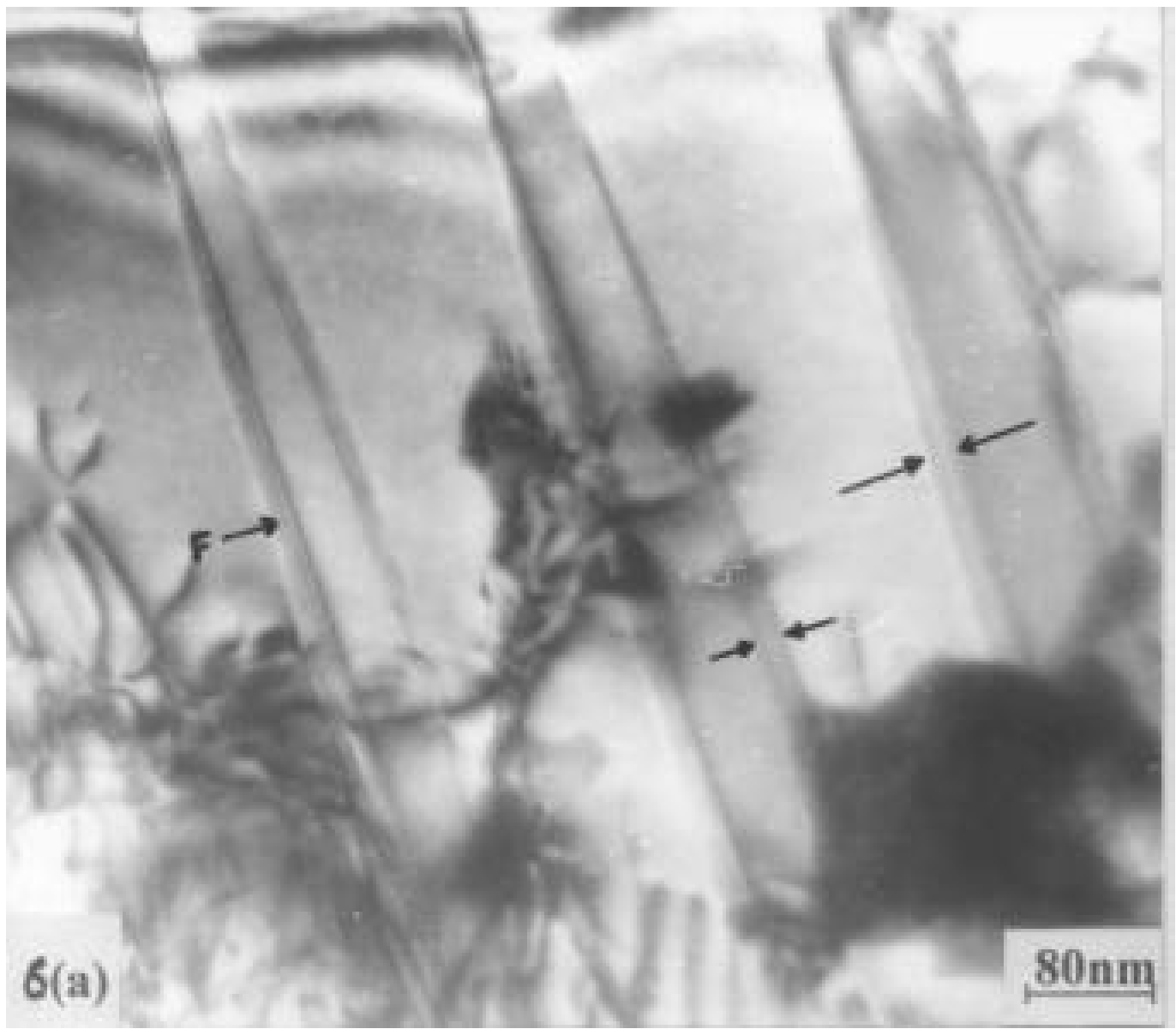

Fig. 6(b) Giri et al.

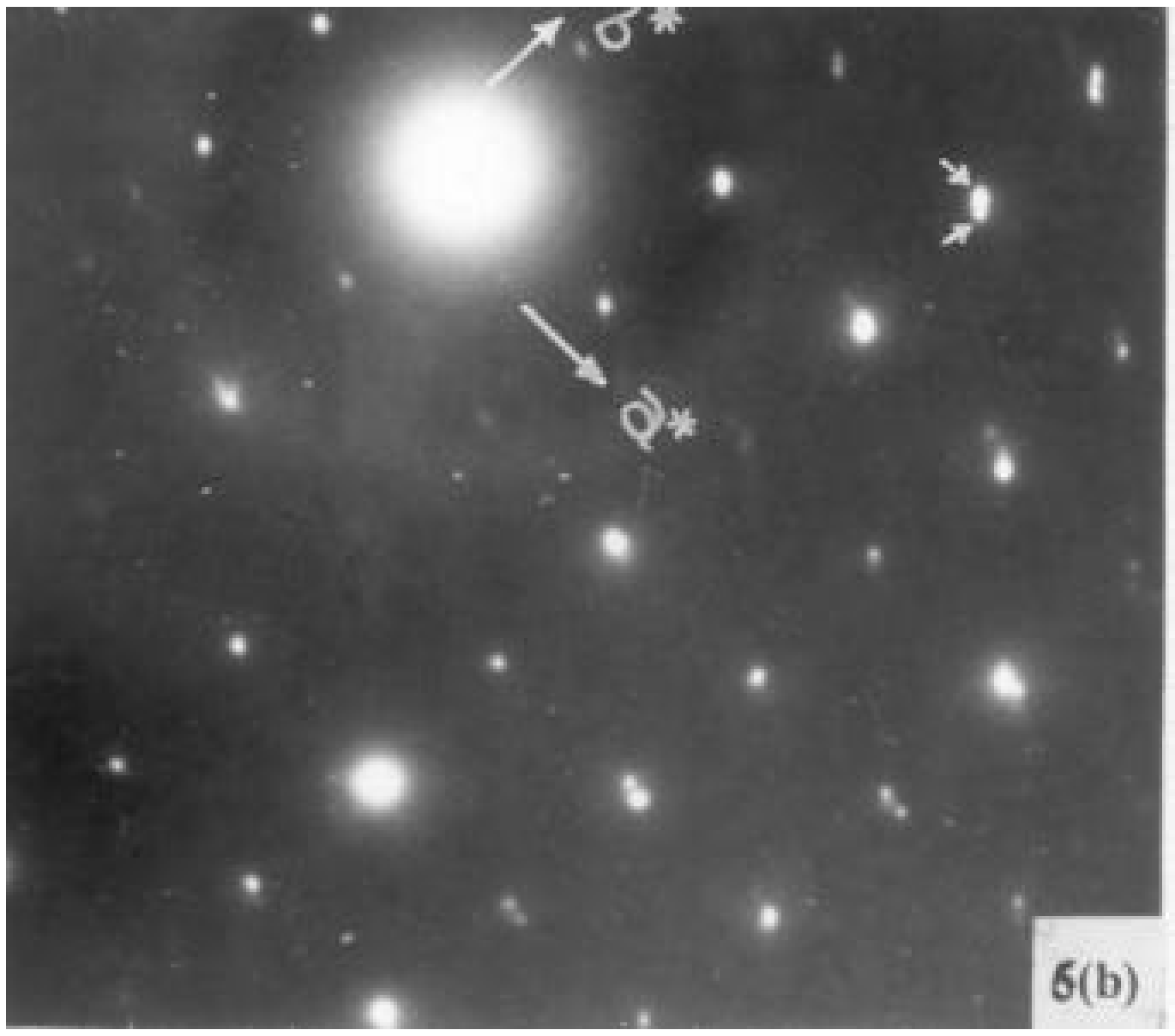

Fig. 6(c) Giri et al.

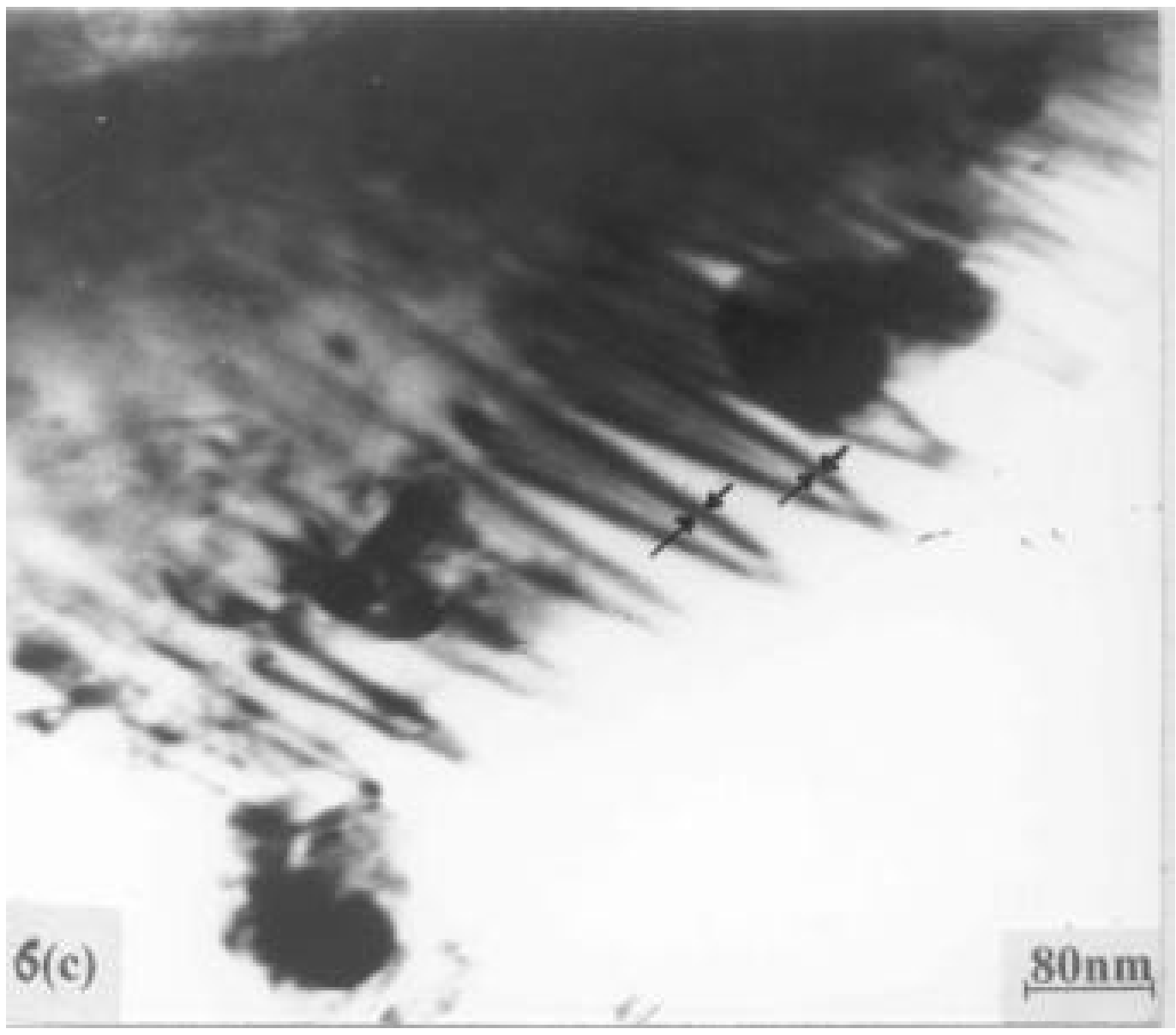

Fig. 6(d) Giri et al.

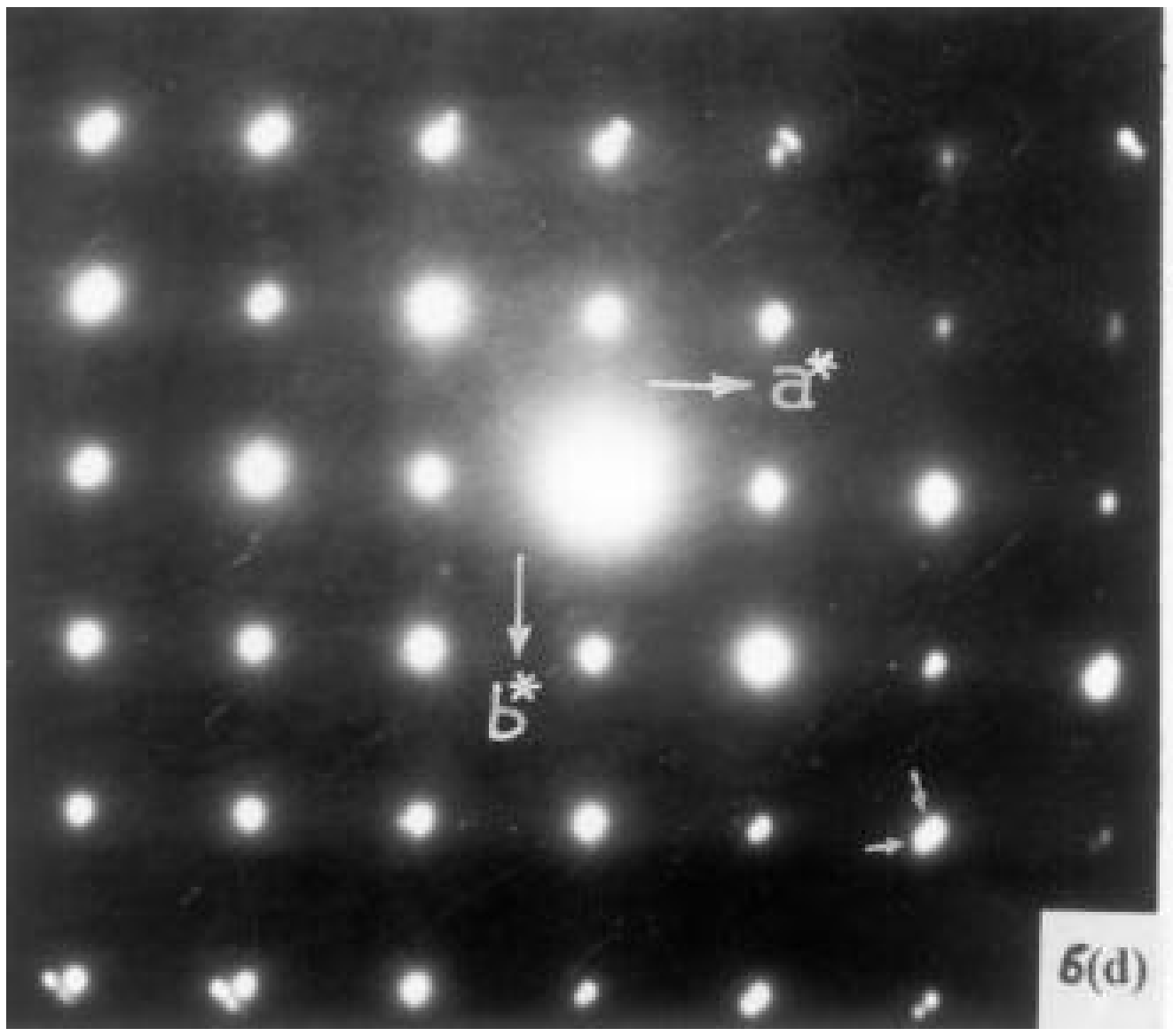

Fig. 6(e) Giri et al.

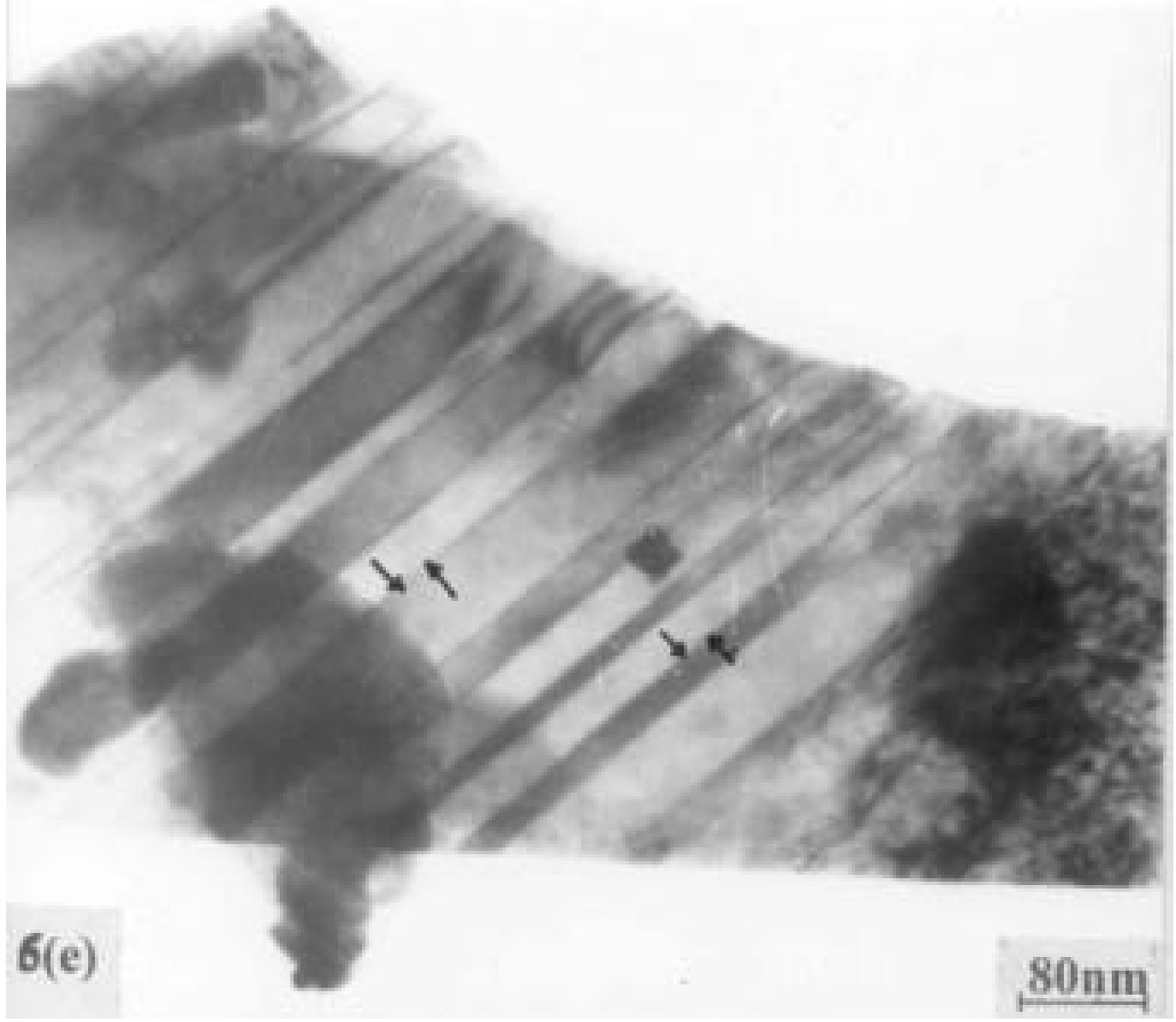

Fig. 6(f) Giri et al.

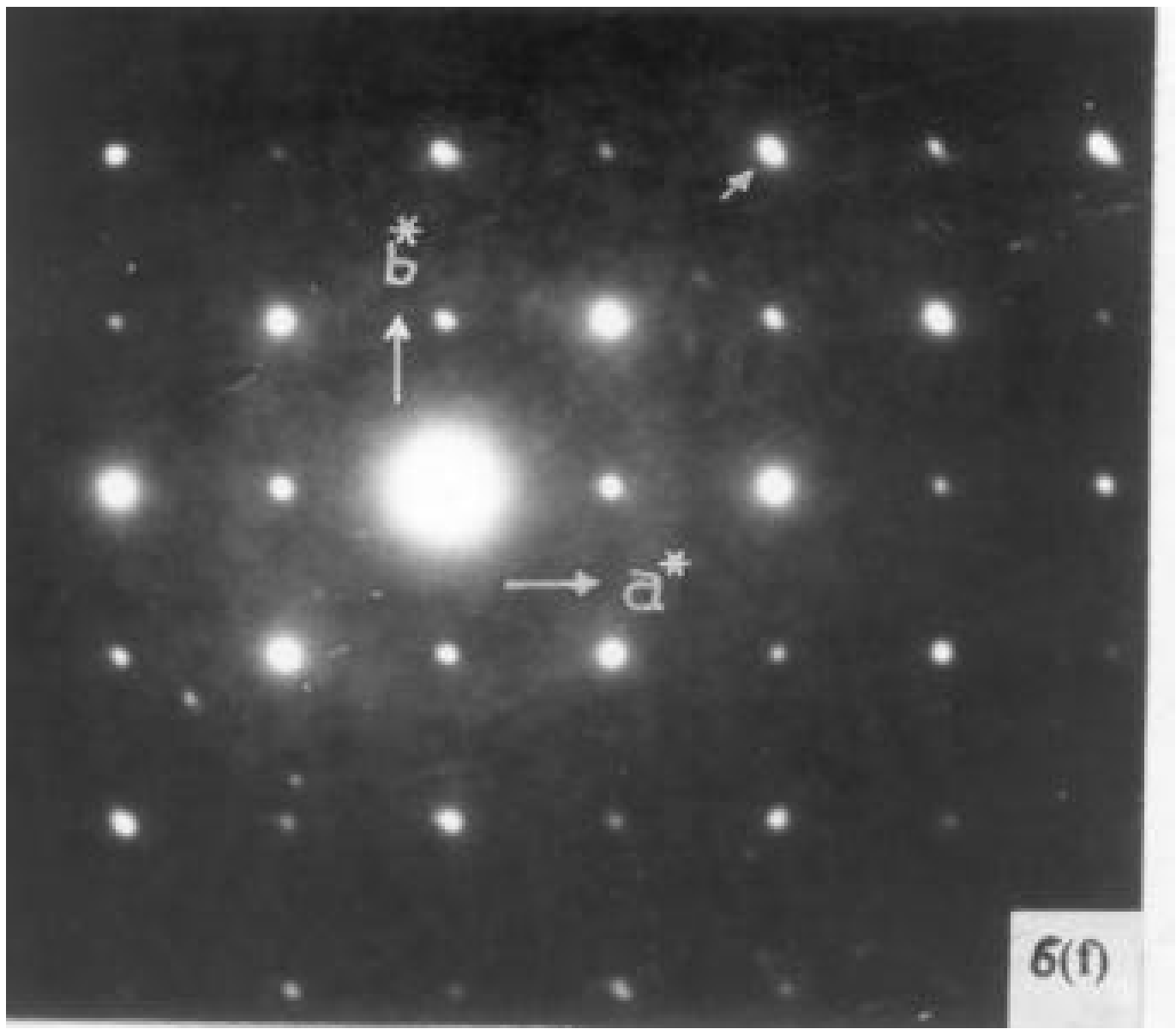